\documentclass[a4paper,12pt]{article}

\usepackage[T2A]{fontenc}
\usepackage{amsmath,amssymb}
\usepackage{cite}
\usepackage[unicode,linktocpage=true,plainpages=false,pdfpagelabels=false]{hyperref}

\newcommand{\dd}{\partial}
\newcommand{\de}{\delta}
\newcommand{\al}{\alpha}
\newcommand{\be}{\beta}
\newcommand{\ka}{\varkappa}
\newcommand{\la}{\lambda}
\newcommand{\pp}{\varphi}
\newcommand{\ep}{\varepsilon}
\newcommand{\ga}{\gamma}
\newcommand{\Ga}{\Gamma}
\newcommand{\simup}[2]{\left\{#1\right\}^{#2}}
\newcommand{\simdown}[2]{\left\{#1\right\}_{#2}}
\newcommand{\antisimup}[2]{\left[#1\right]^{#2}}

\begin{document}

\title{Energy-momentum pseudotensor and superpotential\\
for generally covariant theories of gravity of general form
}

\author{R.~V.~Ilin\thanks{E-mail: rireilin@gmail.com},
S.~A.~Paston\thanks{E-mail: pastonsergey@gmail.com}\\
{\it Saint Petersburg State University, Saint Petersburg, Russia}
}
\date{\vskip 15mm}
\maketitle

\begin{abstract}
The current paper is devoted to the investigation of the general form
of the energy-momentum pseudotensor (pEMT) and the corresponding
superpotential for the wide class of theories. The only requirement
for such a theory is the general covariance of the action without any
restrictions on the order of derivatives of the independent variables
in it or their transformation laws. As a result of the generalized
Noether procedure, we obtain a recurrent chain of the equations, which
allows one to express canonical pEMT as a divergence of the
superpotential. The explicit expression for this superpotential is
also given. We discuss the structure of the obtained expressions and
the conditions for the derived pEMT conservation laws to be satisfied
independently (fully or partially) of the equations of motion.
Deformations of the superpotential form for theories with a change of
the independent variables in the action are also considered. We apply
these results to some interesting particular cases: General Relativity
and its modifications, particularly mimetic gravity and
Regge-Teitelboim embedding gravity.
\end{abstract}

\newpage

\section{Introduction}
{One of the traditional issues of General Relativity (GR) is the correct definition of the conserved quantities related to the space-time symmetries, especially the energy. This problem is not special for GR and in general remains for all generally covariant theories. One may point out two different origins of this problem. First of all, the corresponding densities of the integrals of motion, for example, EMT, are not tensors under the diffeomorphisms (for this reason EMT is often called pseudo-tensor (pEMT)). Moreover, it is always possible to choose the specific reference frame where at any given point the densities vanish. As a result, the total energy-momentum density is not well-defined. This observation was done back at the dawn of the GR (for historical review see \cite{Baryshev, PetrovLompayTekinKopeikin}). Despite the existence of the "covariantization" methods for such densities (see, for example, \cite{linden-bell, BabakGrischukEnergy, statja25}), there is no widely accepted solution for this "non-localizability" problem.}
{Perhaps, the main reason for this is the existence of the newer questions, which have appeared
during the historical development of the GR extensions. The correct description of the dark matter and dark energy,
as well as the inflation theory, at some point, have become central in the context of the topic.} The problem of non-localizability
has remained as a necessary evil, which, however, does not preclude using non-localizable superpotentials for the analysis of the several
astrophysical problems \cite{ThorneHartleTidal,FavataTidalHeat,PatriciaTidalHeating}.

{Aside from the non-localizability problem, there is another obstacle to giving proper definitions for the energy and other the space-time integrals of motion. Namely, there are infinitely many conserved pseudo-tensor currents, which correspond to the same integral of motion. Within the classical set-up, the solution usually comes down to the choice of the number of criteria, which fix the current form \cite{GoldbergConservation, BergmanConserv, Moller, Komar}. This situation is particularly interesting in the quantum theory. As it is discussed, for example, in \cite{AveryNoetherWard}, only a certain currents lead to the correct Ward identities. The currents obtained from standard Noether theorem \cite{noether} are suitable for such purposes. This fact opens new possibilities for the investigations of these quantum identities in the wide class of theories with help of well-known classical results.}

{In the current paper, we study the generalized Noether procedure to obtain explicit formulae for pEMT and the corresponding superpotential for the maximally general theories of gravity with diffeomorphism-invariant action.}
In most GR modifications, the form of the independent variables and their transformation laws are often heavily constrained.
Usually, the maximal tensor rank of these independent variables is equal to two, as well as the maximal order of derivatives in the action.
However, some models violate these restrictions, for example,
such as \cite{mukhanov, InfiniteDerivative, HorndeskiMimetic}.
{To cover most general case,} we impose only one restriction on the action: its gravitational and matter parts
must be invariant under the diffeomorphisms separately.
There are no additional restrictions on the structure of transformation laws for the independent gravitational variables,
or on the maximal (finite) order of their derivatives in the action.

In the section~\ref{GeneralSuperpotentialSec} we generalize the Noether procedure {for such large class}
by using the analysis, that is
similar to the one developed in the
{ \cite{julia}. It is worth noting, that the analysis made in \cite{julia} is limited to the most common case of action with no more than first derivatives of the fields with simple transformation law with respect to the gauge symmetry.}
As a result, we obtain an expression of the pEMT through superpotential and find its {explicit} form.
Despite the lack of the {antisymmetricity} for the superpotential, we show that the conserving energy-momentum
vector can always be expressed through the integral over an infinitely remote surface. {The topology of the space-time in this case is assumed trivial.}
It is worth noting that the general
{properties of}
the superpotential in the gauge theories, particularly for the superpotential that corresponds to pEMT,
was often discussed in the past (see, for example, \cite{BarnichGeneralForm}).
However, in {the current} paper
{we use a much simpler analysis, which allows us to find an explicit expression for the superpotential.}

The relations obtained in the section~\ref{GeneralSuperpotentialSec} are used in section~\ref{SuperpotentialStructureSec}
to perform the analysis of the superpotential structure.
In the section~\ref{ChangeOfVariablesSection} we briefly discuss how the integrals of motion and the related quantities change
in the theories with the smooth change of independent variables in action.
{The first half of the section~\ref{primeri} is devoted to applying the results obtained in the previous sections for the several well-known particular cases: General Relativity and Palatini formalism}. {In the second half of this section we compute new superpotentials for the two modifications of gravity: theories with the disformal change \cite{Bekenstein1993} of variables in Einstein-Hilbert action, including mimetic gravity, and lastly the Regge-Teitelboim theory\cite{regge}.}

\section{General form of the superpotential}\label{GeneralSuperpotentialSec}
Consider the following theory
\begin{equation}
    S\left[y_a,\pp_A\right] = \int dx\, L_{gr}(y_a,..,\dd_{\al_1..\al_{N_1}}y_a)+\int dx\, L_{m}(\pp_A,..,\dd_{\be_1..\be_{N_2}}\pp_A, y_a,..,\dd_{\al_1..\al_{N_3}}y_a),
    \label{GeneralLagrangian}
\end{equation}
where the first term is the gravitational action $S_{gr}$, and the second one is the matter action $S_m$.
The number of the spacetime dimensions does not play any significant role in all following discussions, so it can be considered
arbitrary.
We will use the notation $\dd_{\al_k..\al_i}$ for $\dd_{\al_k}..\dd_{\al_i}$ if $i\geq k$; if $i = k-1$ we will assume that
$\dd_{\al_k..\al_i} = 1$. For all other cases $\dd_{\al_k..\al_i} = 0$.
Quantities $y_a$ denote the independent variables describing the gravitational field, and $\pp_A$ are the matter fields.
{For the rest of the work we will use the following} decomposition for
the transformation law for $y_a$ under infinitesimal diffeomorphisms $x^{\mu}\rightarrow x^{\mu}+\xi^{\mu}$:
\begin{equation}
    \de y_a(x) \equiv y'_a(x)-y_a(x) = -\sum_{i=0}^{M}H_{(i)}{}_{a\rho}{}^{\al_1..\al_i}\dd_{\al_1..\al_i}\xi^{\rho},
    \label{gravityVariablesVariation}
\end{equation}
where $H_{(i)a\rho}{}^{\al_1..\al_i}$ are smooth functions of $y_a$, which are assumed symmetric over indices $\al_1..\al_i$. {Note, that the gravitational variables $y_a$ may not be tensors}.

Index $a$ enumerates variables in the set of all independent gravitational variables and, in particular, may be either a spacetime
index or index of some inner symmetry groups.
Similarly, multi-index $A$ for the matter fields $\pp_A$ may contain both spacetime indices and indices for inner symmetries.
For further analysis, the latter is not significant, so we will omit it. {In the sake of simplicity} we also consider matter fields to be tensors of rank $U$, so their variations is assumed to be the following:
\begin{equation}
    \de\pp_A = -\xi^{\al}\dd_{\al}\pp_A - \sum_{i = 1}^U\dd_{\mu_i}\xi^{\rho}\pp_{(\rho\hat{A})_i},
    \label{matterVariablesVariation}
\end{equation}
where
\begin{equation}
\pp_A \equiv \pp_{\mu_1..\mu_U},\qquad
\pp_{(\rho\hat{A})_i}\equiv\pp_{\mu_1..\mu_{i-1}\rho\mu_{i+1}..\mu_{U}}.
\end{equation}
The parameters $N_1,N_2,N_3$ define the maximal order of the derivatives in the action \eqref{GeneralLagrangian}.
{The number $M$ corresponds to the maximal order of the infinitesimal diffeomorphisms parameter $\xi^{\mu}$ in the transformation law
of the gravitational variables \eqref{gravityVariablesVariation}}.
One may choose $\pp_A$ to be the spinor field in order to describe fermions.
{This} choice requires the matter fields $\pp_A$ to form a spinor representation of the
certain inner gauge group (for 4-dimensional spacetime, the proper choice is the Lorentz group $SO(1,3)$).
As was already mentioned, the inner symmetries are irrelevant in the current analysis; hence the introduction of spinors does not affect the results much.

We require that the gravitational and the matter parts of the action \eqref{GeneralLagrangian} must be {invariant under diffeomorphisms} separately.
In particular, it means that the total action $S$ is invariant under certain global transformations, which, due to the Noether
theorem, lead to the on-shell local conservation laws (continuity equations).
In order to study these quantities, we follow the route presented in \cite{statja53,julia,GoMa}.
We consider an infinitesimal diffeomorphism $x^{\mu}\rightarrow x^{\mu}+\xi^{\mu}(x)$ and by using the arbitrariness of the
vector $\xi^{\mu}$ we will obtain a recurrent chain of identities. {As we will see, it allows one to restrict the form of the canonical pEMT and energy-momentum vector.}

The general covariance of the action implies that the Lagrangian must {the scalar density under the infinitesimal coordinate transformation $x^{\mu}\rightarrow x^{\mu}+\xi^{\mu}$. This leads to the well-known formula}:
\begin{equation}
    \de L = -\dd_{\al}\left(\xi^{\mu}L\right).
    \label{LagrangianTransfLaw}
\end{equation}
{The} statement is also true for $L_{gr}$ and $L_m$, but {we postpone them until} the section~\ref{SuperpotentialStructureSec}.
Note that l.h.s. of the \eqref{LagrangianTransfLaw} can be written in terms of the variations of the independent variables {$\de y_a$ and $\de\pp_A$} in the form:
\begin{equation}
    \de L = \sum_{i = 0}^{\max(N_1,N_3)}\frac{\dd L}{\dd\dd_{\al_1...\al_i}y_a}\dd_{\al_1..\al_i}\de y_a +\sum_{i = 0}^{N_2}\frac{\dd L}{\dd\dd_{\al_1...\al_i}\pp_A}\dd_{\al_1...\al_i}\de\pp_A.
    \label{LagrangianVariationDecomposition}
\end{equation}
We can express the terms with $i = 0$ through $\de S/\de y_a$,  $\de S/\de\pp_A$ and use the product rule:
\begin{equation}
\begin{split}
Y^{\al_1\dots\al_j}\dd_{\al_1..\al_j}Q
+(-1)^{j+1}(\dd_{\al_1..\al_j}Y^{\al_1..\al_j})Q=&\\
=\dd_{\rho}\Bigg[\sum_{i=0}^{j-1}(-1)^{i+j+1}(\dd_{\al_1..\al_{j-i-1}}Y^{\rho\al_1..\al_{j-1}})&(\dd_{\al_{j-i}..\al_{j-1}} Q)\Bigg],
    \label{HigherOrderFullDerivativeRelation}
\end{split}
\end{equation}
which is valid for the arbitrary symmetric tensor $Y^{\al_1...\al_j}$, to obtain the following relation:
\begin{align}
    \dd_{\rho}J^{\rho}=-\frac{\de S}{\de y_a}\de y_a-\frac{\de S}{\de \pp_A}\de \pp_A,
    \label{OffShellTotalNoetherIdentity}
\end{align}
where
\begin{align}
\nonumber
    J^{\rho} \equiv \xi^{\rho}L+&\sum_{j=1}^{\max(N_1,N_3)}\sum_{i = 0}^{j-1}(-1)^{i+j+1}(\dd_{\al_1..\al_{j-i-1}}Z^{a\rho\al_1..\al_{j-1}})(\dd_{\al_{j-i}..\al_{j-1}}\de y_a)+\\
    &+\sum_{j=1}^{N_2}\sum_{i = 0}^{j-1}(-1)^{i+j+1}(\dd_{\al_1..\al_{j-i-1}}N^{A\rho\al_1..\al_{j-1}})(\dd_{\al_{j-i}..\al_{j-1}}\de \pp_A).
\label{JDefinition}
\end{align}
Here we define
$Z^{a\rho\al_1..\al_{j}}$ and $N^{A\rho\al_1..\al_{j-1}}$
as in the following:
\begin{align}
Z^{a\al_1..\al_j} \equiv \frac{\dd L}{\dd\dd_{\al_1..\al_j}y_a},\;\;\;\; N^{A\al_1..\al_j} \equiv \frac{\dd L}{\dd\dd_{\al_1..\al_j}\pp_A}.
\label{ZetNDefinitions}
\end{align}

One can use the transformation laws \eqref{gravityVariablesVariation} and \eqref{matterVariablesVariation}
in order to write \eqref{JDefinition} as the linear combination of $\xi^{\al}$ and its derivatives:
\begin{align}
J^{\rho}=-
\sum_{k=0}^{N_{d}}K_{(k)}{}_{\al}{}^{\rho\al_1..\al_k}\dd_{\al_1..\al_k}\xi^{\al},
\label{KDecomposition2}
\end{align}
where $N_{d}\equiv\max(N_1 + M - 1, N_3 + M - 1,N_2)$.
The formulae for the coefficients $K_{(k)}{}_{\al}{}^{\rho\al_1..\al_k}$ are the following (they may be found by rewriting
\eqref{JDefinition} in the form \eqref{KDecomposition2}):
\begin{align}
\nonumber
    &K_{(0)}{}_{\al}{}^{\rho} ={} \sum_{j=1}^{\max(N_1,N_3)}\sum_{i = 0}^{j-1}(-1)^i(\dd_{\al_{j-i}..\al_{j-1}}Z^{a\rho\al_1..\al_{j-1}})(\dd_{\al_{1}..\al_{j-i-1}}H_{(0)}{}_{a\al}) +\\&\hspace{2.5cm}+\sum_{j=1}^{N_2}\sum_{i = 0}^{j-1}(-1)^i(\dd_{\al_{j-i}..\al_{j-1}}N^{A\rho\al_1..\al_{j-1}})(\dd_{\al_1..\al_{j-i-1}\al}\pp_A) - \de^{\rho}_{\al}L,\\
\nonumber
    &K_{(k)}{}_{\al}{}^{\rho\al_1..\al_k} ={} \theta\left(\max(N_1,N_3)+M-k-\frac{1}{2}\right)\sum_{l = \max(k-\max(N_1,N_3)+1,0)}^{\min(k,M)}\Omega_{(1)}{}^{\rho}{}_{\al}{}^{\al_1..\al_k}(k-l,l) +\\
    &\hspace{2.5cm}+\theta\left(N_2-k+\frac{1}{2}\right)\left(\Omega_{(2)}{}^{\rho}{}_{\al}{}^{\al_1..\al_k}(k) + \Omega_{(3)}{}^{\rho\al_k}{}_{\al}{}^{\al_1..\al_{k-1}}(k-1)\right),
    \quad 1\leq k \leq N_d,\label{sp1}
\end{align}
where
\begin{align}
\nonumber
    \Omega_{(1)}{}^{\rho}{}_{\mu}{}^{\al_1..\al_m\be_1..\be_l}(m,l) \equiv& \Bigg\{\sum_{j=m+1}^{\max(N_1,N_3)}\sum_{i = 0}^{j-m-1}(-1)^iC^m_{j-i-1}(\dd_{\al_{j-i}..\al_{j-1}}Z^{a\rho\al_1..\al_{j-1}})\times\\&\hspace{4.5cm}\times\dd_{\al_{m+1}..\al_{j-i-1}}H_{(l)}{}_{a\mu}{}^{\be_1..\be_l}\Bigg\}^{\al_1..\al_m\be_1..\be_l},
    \label{OmegaFirstDefinition}\\
\nonumber
    \Omega_{(2)}{}^{\rho}{}_{\mu}{}^{\al_1..\al_{m}}(m) \equiv& \Bigg\{\sum_{j=m+1}^{N_2}\sum_{i = 0}^{j-m-1}(-1)^iC^m_{j-i-1}(\dd_{\al_{j-i}..\al_{j-1}}N^{A\rho\al_1..\al_{j-1}})\times\\&\hspace{6.5cm}\times\dd_{\al_{m+1}..\al_{j-i-1}\mu}\pp_A\Bigg\}^{\al_1..\al_m},
    \label{OmegaSecondDefinition}
\end{align}

\begin{align}
\nonumber
    \Omega_{(3)}{}^{\rho\sigma}{}_{\mu}{}^{\al_1..\al_{m}}(m) \equiv&
    \Bigg\{\sum_{j=m+1}^{N_2}\sum_{i = 0}^{j-m-1}(-1)^iC^m_{j-i-1}\times\\
    &\hspace{1cm}\times\left[\sum_{k = 1}^U(\dd_{\al_{j-i}..\al_{j-1}}N^{(\sigma A)_k\rho\al_1..\al_{j-1}})\dd_{\al_{m+1}..\al_{j-i-1}}\pp_{(\mu A)_k}\right]\Bigg\}^{\al_1..\al_m\sigma},
    \label{OmegaThirdDefinition}
\end{align}
and $\{..\}^{\al_1..\al_k}$ denotes symmetrization with respect to the indices $\al_1..\al_k$ -- sum of the expression in the brackets
over all permutations of the mentioned indices.
Note that $K_{(k)}{}_{\al}{}^{\rho\al_1..\al_k}$ are fully symmetric with respect to the indices $\al_1..\al_k$ by construction.

By using the identities \eqref{KDecomposition2} one can write \eqref{OffShellTotalNoetherIdentity} in the more convenient form:
\begin{align}
    \dd_{\rho}\left(\sum_{k=0}^{N_{d}}K_{(k)}{}_{\al}{}^{\rho\al_1..\al_k}\dd_{\al_1..\al_k}\xi^{\al}\right) = \frac{\de S}{\de y_a}\de y_a+\frac{\de S}{\de \pp_A}\de \pp_A.
\label{KDecomposition}
\end{align}
The r.h.s of the relations \eqref{KDecomposition} vanishes on-shell, so in this case these relations describe local conservation law, that
is satisfied for any choice of $\xi^{\al}(x)$.
If one takes $\xi^{\al} = const$, the identity \eqref{KDecomposition} gives an expected on-shell equation:
\begin{equation}
    \dd_{\rho}K_{(0)}{}_{\al}{}^{\rho} = 0.
    \label{KZeroConservation}
\end{equation}
As the reasoning for $\xi^{\al} = const$ is exactly the first Noether theorem for the spacetime translation invariance, we will use the proper notation:
\begin{equation}
    K_{(0)}{}_{\al}{}^{\rho} = \mathcal{T}^{\rho}{}_{\al},
    \label{KZeroEMTEquivalence}
\end{equation}
where $\mathcal{T}^{\rho}{}_{\al}$ is canonical pEMT.

Let us return to the general choice of the $\xi^{\al}(x)$.
As it is completely arbitrary, one can independently choose its value and the value of all its derivatives $\dd_{\al_1..\al_k}\xi^{\mu}$ at
any given point.
Hence it is possible to rewrite the \eqref{KDecomposition} as the recurrent chain of identities
\begin{align}
    &\dd_{\rho}K_{(0)}{}_{\al}{}^{\rho} = -\frac{\de S}{\de y_a}H_{(0)a\al}-\frac{\de S}{\de\pp_A}\dd_{\al}\pp_A,\label{KDecompositionReccurentChain0}\\
\nonumber
    &\dd_{\rho}K_{(k)}{}_{\al}{}^{\rho\al_1..\al_k}+\frac{1}{k!}\left\{K_{(k-1)}{}_{\al}{}^{\al_1..\al_k}\right\}^{\al_1..\al_k}=\\
    &\hspace{2cm}=-\frac{\de S}{\de y_a}H_{(k)a\al}{}^{\al_1..\al_k}\theta\left(M-k+\frac{1}{2}\right)-\de_{1k}\frac{\de S}{\de \pp_A}\sum_{i = 1}^U\de^{\al_1}_{\mu_i}\pp_{(\al A)_i},
    \hspace{1cm} 1 \le k \le N_{d},
    \label{KDecompositionReccurentChain1}\\ &\left\{K_{(N_{d})\al}{}^{\al_1...\al_{N_{d}+1}}\right\}^{\al_1...\al_{N_{d}+1}} = 0. \hspace{3cm} &
\label{KDecompositionReccurentChain2}
\end{align}
The analogous recurrent chain arises in the \cite{statja53} in the context of generalized Belinfante relation.
{One may refine the recurrent chain \eqref{KDecompositionReccurentChain0}-\eqref{KDecompositionReccurentChain2} by following the procedure described in \cite{statja53} on-shell in order to obtain the general form of superpotential for pEMT of theory \eqref{GeneralLagrangian}. The resulting superpotential is not nescessarily antisymmetric. Though it prevents one from writing a simple expression of the energy-momentum vector through the surface integral, we will show, that even without antisymmetricity it may be done explicitly.}

In the rest of this section, we assume that all equations of motion are satisfied.
The equation with $k = 1$ from \eqref{KDecompositionReccurentChain1} has the following form (here we used \eqref{KZeroEMTEquivalence}):
\begin{equation}
    \mathcal{T}^{\rho}{}_{\al} = - \dd_{\be}K_{(1)\al}{}^{\be\rho}.
    \label{SuperpotentialAnalysisResult}
\end{equation}
This equation expresses the canonical pEMT as the full divergence of a certain quantity, so its reasonable to
think of $K_{(1)\al}{}^{\be\rho}$ as the superpotential (up to a possible sign convention).
However, the definition of the superpotential, i.e., the property
\begin{equation}
    \dd_{\be\rho}K_{(1)\al}{}^{\be\rho} = 0
    \label{KsuperpotentialProperty}
\end{equation}
should be proven first.
To see it, one should apply the operator $\dd_{\al_1..\al_k}$ to the
\eqref{KDecompositionReccurentChain1}, the resulting identity will be the following:
\begin{align}
    &\dd_{\rho\al_1..\al_k}K_{(k)}{}_{\al}{}^{\rho\al_1..\al_k} = - \dd_{\al_1..\al_k}K_{(k-1)}{}_{\al}{}^{\al_1..\al_k},
\end{align}
and also by applying $\dd_{\al_1..\al_{N_d+1}}$ to the \eqref{KDecompositionReccurentChain2}, one can obtain the similar formula:
\begin{align}
    &\dd_{\al_1..\al_{N_d+1}}K_{(N_d)}{}_{\al}{}^{\al_1..\al_{N_d+1}} = 0.
\end{align}
From these equations it is clear that
\begin{align}
    &\dd_{\al_1..\al_{k+1}}K_{(k)}{}_{\al}{}^{\al_1..\al_{k+1}} =0,\qquad 0 \le k \le N_{d},
\end{align}
from which the superpotentiality \eqref{KsuperpotentialProperty} of $K_{(1)\al}{}^{\be\rho}$ arises.
By using \eqref{sp1} with $k=1$ we finally obtain the general form of the superpotential $\Psi_{\al}{}^{\be\rho}$:
\begin{align}
&\hspace{6.5cm}\mathcal{T}^{\rho}{}_{\al} = \dd_{\be}\Psi_{\al}{}^{\be\rho},\\
&\Psi_{\al}{}^{\be\rho} =-K_{(1)\al}{}^{\be\rho}=
-\theta\left(\max(N_1,N_3)+M-\frac{3}{2}\right)\sum_{l = \max(2-\max(N_1,N_3),0)}^{\min(1,M)}\Omega_{(1)}{}^{\be}{}_{\al}{}^{\rho}(1-l,l)\nonumber\\
&\hspace{2.5cm}-
\theta\left(N_2-\frac{1}{2}\right)\left(\Omega_{(2)}{}^{\beta}{}_{\al}{}^{\rho}(1) + \Omega_{(3)}{}^{\beta\rho}{}_{\al}(0)\right).
\label{sp2}
\end{align}

For particular cases of \eqref{GeneralLagrangian} with $N_1,N_2,N_3\leq 2$
instead of $\Psi_{\al}{}^{\be\rho}$ one can \cite{PetrovLompayTekinKopeikin} use the following superpotential:
\begin{equation}
B^{\be\rho}{}_{\al}= -K_{(1)\al}{}^{\be\rho} + \dd_{\gamma}L_{\al}{}^{\be\gamma\rho}, \qquad
B^{\be\rho}{}_{\al} = -B^{\rho\be}{}_{\al},
\label{AntisymmetricSuperpotentialDefinition}
\end{equation}
where $L_{\al}{}^{\be\gamma\rho}$ has the property:
\begin{equation}
\dd_{\be\gamma}L_{\al}{}^{\be\gamma\rho} = 0.
\label{sp5}
\end{equation}
Then the canonical pEMT can be written in the equivalent form in terms of the described antisymmetric superpotential
$B^{\be\rho}{}_{\al}$:
\begin{equation}
\mathcal{T}^{\rho}{}_{\al} = \dd_{\be}B^{\be\rho}{}_{\al}.
\end{equation}
Though antisymmetricity of the superpotential provides a simple expression for the energy-momentum in terms of the surface integral,
it can be difficult to construct $L_{\al}{}^{\be\gamma\rho}$ explicitly in the general case.
However, {as it was mentioned above,} there is another option to achieve the same goal.
In order to do it, let us consider \eqref{KDecompositionReccurentChain1} with $\al_1 =\al_2=\dots=\al_k = 0$:
\begin{equation}
    K_{(k-1)}{}_{\al}{}^{0\dots0} = - \dd_0K_{(k)}{}_{\al}{}^{ 0\dots0}-\dd_n K_{(k)}{}_{\al}{}^{ n0\dots0},
    \label{KChainWithZeroIndices}
\end{equation}
where index $n$ corresponds to the spatial dimensions.
By applying the operator $(-\dd_0)^{i-2}$ to this equation, summing the result from $k = 2$ to $N_d$ and taking into
account the \eqref{KDecompositionReccurentChain2}
\begin{equation}
K_{(N_d)}{}_{\al}{}^{0\dots0} = 0,
\end{equation}
it can be shown that
\begin{equation}
    K_{(1)}{}_{\al}{}^{00} = -\dd_n\sum_{i = 2}^{N_d}(-\dd_0)^{i-2}K_{(i)}{}_{\al}{}^{n0\dots0}.
    \label{eq:K_1WithZeros}
\end{equation}
Equation \eqref{eq:K_1WithZeros} with assumptions of the trivial spacetime topology and the proper field asymptotics on
the spatial infinity leads to the expression for the conserving energy-momentum vector based on \eqref{SuperpotentialAnalysisResult}:
\begin{equation}
P_{\al} \equiv \int\limits_{t={const}} d\vec{x}\; \mathcal{T}^0{}_{\al} =
\lim_{R\rightarrow\infty}\int\limits_{S_R} dS_n\left[-K_{(1)}{}_{\al}{}^{n0}-\sum_{i = 2}^{N_{d}}(-\dd_0)^{i-1}K_{(i)}{}_{\al}{}^{n0...0}\right].
\label{Energy-MomentumConservation}
\end{equation}
{Here} $S_R$ is a sphere of radius $R$ in the hyperspace $t = const$.

\section{Properties of gravitational and matter contributions into superpotential}\label{SuperpotentialStructureSec}
One of the requirements imposed on the action \eqref{GeneralLagrangian} was the separate general covariance of both gravitational and matter
{parts}.
It allows one to obtain the recurrent chains similar to the chain \eqref{KZeroConservation},\eqref{KDecompositionReccurentChain1},\eqref{KDecompositionReccurentChain2}
for each of these action contributions.
In this section, we show that such chains allow one to lend certain physical sense to the gravitational and matter contributions into the
superpotential \eqref{SuperpotentialAnalysisResult}.

Because of the general covariance of the $S_m$, the lagrangian $L_m$ is {obviously the scalar density. It means} we can follow
the same procedure given by the identities \eqref{LagrangianTransfLaw}-\eqref{OffShellTotalNoetherIdentity}, and derive the analogue for the
\eqref{OffShellTotalNoetherIdentity}:
\begin{equation}
    \dd_{\rho}\big[J^{\rho}\big]_{m}=-\frac{\de S_m}{\de y_a}\de y_a-\frac{\de S_m}{\de \pp_A}\de\pp_A,
    \label{NoetherIdentityForMatterAction}
\end{equation}
where $J^{\rho}$ is defined in \eqref{JDefinition}, and $[..]_m$ denotes the contribution from $L_m$ extracted from the quantity in brackets.

For further analysis of this identity, it is helpful to introduce the additional identities that also follow from the general covariance.
Consider the infinitesimal change $x^{\mu}\rightarrow x^{\mu} + \xi^{\mu}(x)$ with $\xi^{\mu}$ having a compact support.
{The general covariance} of the total action {implies} that:
\begin{equation}
    \de S = \int dx\left( \frac{\de S}{\de y_a}\de y_a + \frac{\de S}{\de\pp_A}\de\pp_A \right) = 0,
    \label{matterActionVariation}
\end{equation}
where $\de y_a$ and $\de\pp_A$ are defined in \eqref{gravityVariablesVariation} and \eqref{matterVariablesVariation}.
Integration by parts can remove all the derivatives from $\xi^{\mu}$ {because we assumed it has compact support. Then it is not hard to check, that} the following identities {are satisfied}:
\begin{equation}
    \frac{\de S}{\de \pp_A}\dd_{\rho}\pp_A-\sum_{i=1}^U\dd_{\mu_i}\left(\frac{\de S}{\de \pp_A}\pp_{(\rho\hat{A})_i}\right)+\sum_{i = 0}^M(-1)^i\dd_{\al_1..\al_i}\left(H_{(i)a\rho}{}^{\al_1..\al_i}\frac{\de S}{\de y_a}\right) = 0.
    \label{MetricEMTCovariantConservationGeneralization}
\end{equation}
This relation establishes a useful link between different equations of motion.
The existence of such identities is a well-known fact for any gauge theory and first was pointed by Noether in \cite{noether}.
In literature (see, for example, \cite{BarnichGeneralForm}), it also called the second Noether theorem.

By isolating the term with $i=0$ in the last sum in \eqref{MetricEMTCovariantConservationGeneralization}
and also by changing $S$ to $S_m$ due to the general covariance of the matter action, it can be shown that:
\begin{equation}
    H_{(0)a\rho}\frac{\de S_m}{\de y_a} =\sum_{i=1}^U\dd_{\mu_i}\left(\frac{\de S_m}{\de \pp_A}\pp_{(\rho\hat{A})_i}\right) - \frac{\de S_m}{\de \pp_A}\dd_{\rho}\pp_A - \sum_{i = 1}^M(-1)^i\dd_{\al_1..\al_i}\left(H_{(i)a\rho}{}^{\al_1..\al_i}\frac{\de S_m}{\de y_a}\right).
\label{MetricEMTCovariantConsAlternativeForm}
\end{equation}
In the particular case of GR (when the gravitational independent variable $y_a$ is reduced to $g_{\mu\nu}$ and hence $M = 1$) the identity \eqref{MetricEMTCovariantConsAlternativeForm}
takes the familiar form if we assume that the matter equations are satisfied:
\begin{equation}
    D_{\mu}T^{\mu\nu} = 0,
    \label{MetricEMTCovariantConservation}
\end{equation}
where
\begin{equation}
T^{\mu\nu} \equiv -\frac{2}{\sqrt{-g}}\frac{\de S_m}{\de g_{\mu\nu}}
\label{MetricEMTDefinition}
\end{equation}
is the metric (Hilbert) EMT of the matter.
The identity \eqref{MetricEMTCovariantConsAlternativeForm} arising from the second Noether theorem may be considered independently
from the chain equations \eqref{KDecompositionReccurentChain0}-\eqref{KDecompositionReccurentChain2}.
As one shall see below, it can be used to simplify them further.

Now we can use \eqref{gravityVariablesVariation} and \eqref{matterVariablesVariation} in \eqref{NoetherIdentityForMatterAction} and then
use \eqref{MetricEMTCovariantConsAlternativeForm} and \eqref{HigherOrderFullDerivativeRelation} in the result to obtain the following:
\begin{align}
    \nonumber &\dd_{\rho}\left(\big[J^{\rho}\big]_{m}-\sum_{j=1}^M\sum_{i = 0}^{j-1}(-1)^{i+j+1}\dd_{\al_1..\al_{j-i-1}}
    \left(H_{(j)a\mu}{}^{\rho\al_1..\al_{j-1}}\frac{\de S_m}{\de y_a}\right)\dd_{\al_{j-i}..\al_{j-1}}\xi^{\mu}+\right.\\&\hspace{10cm}-
    \sum_{i=1}^U\frac{\de S_m}{\de \pp_{(\rho\hat{A})_i}}\pp_{(\nu\hat{A})_i}\xi^{\nu}\Bigg)=0.
    \label{MatterFullDivergenceRelation}
\end{align}
As with \eqref{KDecomposition}, {from the arbitrariness of} $\xi^{\mu}$ {immediately follows,} that this identity is equivalent to some recurrent chain of equations. {It is clear, that it} should look similar to {the chain} \eqref{KZeroConservation},\eqref{KDecompositionReccurentChain1},\eqref{KDecompositionReccurentChain2} which arises
from \eqref{KDecomposition} in the previous section.

For the further analysis of the {chain  \eqref{MatterFullDivergenceRelation}}, we will assume that we are working on-shell in the sense of
the matter equations of motion $\de S_m/\de \pp_A=0$.
Then the first two {equations from} the \eqref{MatterFullDivergenceRelation} looks like the following:
\begin{align}
    &\dd_{\rho}\left[\mathcal{T}^{\rho}{}_{\la}-X^{\rho}{}_{\la}\right]_{m}=0,
    \label{MatterHalfIdentity}\\
    \left[\mathcal{T}^{\rho}{}_{\la}\right.&\left.\right]_m=\left[V^{\rho}{}_{\la}\right]_{m} - \dd_{\be}\left[K_{(1)\al}{}^{\be\rho}\right]_m,
    \label{MatterSuperpotentialRelation}
\end{align}
where we used the notation:
\begin{align}
    X^{\rho}{}_{\mu} \equiv &\sum_{j = 1}^M(-1)^{j}\dd_{\al_1..\al_{j-1}}\left(H_{(j)a\mu}{}^{\rho\al_1..\al_{j-1}}\frac{\de S}{\de y_a}\right),
    \label{XQuantityDefinition}\\
    &\hspace{1cm}V^{\rho}{}_{\mu} \equiv - H_{(1)a\mu}{}^{\rho}\frac{\de S}{\de y_a}.\label{sp4}
\end{align}
Obviously, these quantities vanish on the gravitational equations of motion, so we will use them only in the relations that
are satisfied without the appropriate equations of motion.

For the case mentioned above, when the gravitational variable $y_a$ is spacetime metric $g_{\mu\nu}$, we have:
\begin{equation}
H_{(1)\mu\nu\al}{}^{\rho} = \left\{\de^{\rho}_{\mu}g_{\al\nu}\right\}_{\mu\nu},
\end{equation}
from which we have
\begin{equation}
    \left[X^{\rho}{}_{\mu}\right]_m = \left[V^{\rho}{}_{\mu}\right]_m = \sqrt{-g}\,T^{\rho}{}_{\mu}.
\end{equation}
Substituting these formulae in \eqref{MatterSuperpotentialRelation}, it is not hard to derive the generalization of the well-known Belinfante
relation \cite{belifante} in the curved spacetime:
\begin{equation}
    \left[\mathcal{T}^{\rho}{}_{\la}\right]_m = \sqrt{-g}\,T^{\rho}{}_{\la} -\dd_{\be}\left[K_{(1)\la}{}^{\be\rho}\right]_m.
    \label{Belinfante}
\end{equation}

One may follow the logic of the derivation of the \eqref{MatterHalfIdentity} and \eqref{MatterSuperpotentialRelation} and
then obtain the analogous relations for the $S_{gr}$, and hence, the analogous chain of equations. Repeating these first steps, we obtain essentially the same result:
\begin{align}
    &\dd_{\rho}\left[\mathcal{T}^{\rho}{}_{\la}-X^{\rho}{}_{\la}\right]_{gr}=0,
    \label{GravityHalfIdentity}\\
    \left[\mathcal{T}^{\rho}{}_{\la}\right.&\left.\right]_{gr}=\left[V^{\rho}{}_{\la}\right]_{gr} - \dd_{\be}\left[K_{(1)\al}{}^{\be\rho}\right]_{gr}.
    \label{GravitySuperpotentialRelation}
\end{align}
The only difference will be the obvious lack of the matter fields $\pp_A$ in the $S_{gr}$.
Therefore, in contrast to \eqref{MatterHalfIdentity} and \eqref{MatterSuperpotentialRelation}, the relations
\eqref{GravityHalfIdentity} and \eqref{GravitySuperpotentialRelation} are satisfied off-shell
and hence are true identities.

The relations \eqref{MatterSuperpotentialRelation} and \eqref{GravitySuperpotentialRelation} give a useful insight into
the structure of the general superpotential \eqref{sp2}.
If we just sum these relations and require the satisfaction of only the matter equations of motion to be able to use generalized Belinfante
relation \eqref{MatterSuperpotentialRelation}, we obtain the expected relation for the total pEMT:
\begin{equation}
    \mathcal{T}^{\rho}{}_{\la} = -\dd_{\al}K_{(1)}{}_{\la}{}^{\al\rho} + V^{\rho}{}_{\la},
    \label{FullOffShellConservation}
\end{equation}
which obviously coincides with the \eqref{SuperpotentialAnalysisResult} on-shell because $V^{\rho}{}_{\la} = 0$ on the
gravitational equations of motion.
Summing all up, the quantity $K_{(1)}{}_{\la}{}^{\al\rho}$ (and hence the superpotential \eqref{sp2}) can be naturally
decomposed into the two contributions from the matter and gravitational actions.
The first one can be treated (if the matter equations of motion are satisfied) as the generalization of the Belinfante addition (see \eqref{Belinfante}) to
the canonical pEMT{. For the theories with no more than first derivatives in action in the flat limit, it can be expressed in terms of the spin tensor (see, for example,
\cite{belifante,belifante2}). The second term is certain superpotential, that depends only on the
gravitational independent variables and can be reduced to M{\o}ller superpotential \cite{Moller} in the case of GR.}

As it was already mentioned, $V^{\rho}{}_{\la}$ is proportional to the gravitational equations of motion with the coefficient $H_{(1)a\mu}{}^{\rho}$.
{Suppose, that }one field from the set of independent variables $y_a$ does not have first-order derivatives of $\xi^{\mu}$ in the transformation law
\eqref{gravityVariablesVariation} and hence satisfies the condition $H_{(1)a\mu}{}^{\rho} = 0$. {Then $V^{\rho}{}_{\la}$ does not depend on the equation of motion derived from the variation of the \eqref{GeneralLagrangian}
with respect to this field.}
In the case when $H_{(1)a\mu}{}^{\rho} = 0$ is satisfied for all fields, the relation \eqref{FullOffShellConservation} is drastically simplified
(we again assume that the matter equations of motion are satisfied)
\begin{equation}
    \mathcal{T}^{\rho}{}_{\la} = -\dd_{\al}K_{(1)}{}_{\la}{}^{\al\rho}
    \label{ScalarOffShellContinuityEquation}
\end{equation}
without any gravitational equations of motion.

If we additionally cancel the matter contribution in the total action ($S_m = 0$) and impose stronger conditions on the transformation laws:
\begin{equation}
    H_{(i)\mu}{}^{\al_1..\al_i} = 0,\;\;\; i > 0,
\end{equation}
then the r.h.s of the \eqref{KDecompositionReccurentChain1} identically vanishes, and hence the superpotentiality condition
\eqref{KsuperpotentialProperty} is automatically satisfied.
Thus, the formula \eqref{ScalarOffShellContinuityEquation} can be further strengthened:
\begin{equation}
\dd_{\rho}\mathcal{T}^{\rho}{}_{\la} = 0.
\label{ScalarOffShellContinuityEquation2}
\end{equation}
{Like the equation \eqref{ScalarOffShellContinuityEquation}, this relation does not need any gravitational equations of motion to be satisfied.}
This statement may seem strange because the r.h.s. of the \eqref{KDecompositionReccurentChain0} still depends on the term that
is proportional to the equations of motion (note that $H_{(0)}$ is always non-zero).
However, it follows from the relation \eqref{MetricEMTCovariantConservationGeneralization} that \eqref{KDecompositionReccurentChain0}
identically equals to zero for the considered case.

\section{Theories with the change of independent gravitational variables in action} \label{ChangeOfVariablesSection}
In many cases, the consideration of theories of gravity that are modified by the change of independent {variables in the GR action} may be quite useful. In general, this change may contain derivatives of the new independent variables.
Several examples of GR modifications of this kind will be discussed in section~\ref{primeri}. {However}, there are also other frequently discussed theories, for example,
tetrad formulation of GR and theories with nonzero torsion.
In general, by the change of variables we mean the following expression:
\begin{equation}
    y_a = f_a(y'_E,..,\dd_{\al_1..\al_W}y'_E),
    \label{ChangeOfVariables}
\end{equation}
where $y'_E$ are new gravitational independent variables and $f_a$ is a smooth function.
The action of such theory remains unchanged in terms of dependence on the old variables.
As discussed in the works \cite{mukhanov,statja60}, such modification could lead to richer dynamics
in comparison to the original theory.
Namely, a new theory has all the solutions from the original one and also has some extra solutions, that may appear useful
for the explanations of certain observable effects that are absent in the original theory.

Despite the general form of the \eqref{ChangeOfVariables}, it constrains the form of the integrals of motion in the modified theory.
Indeed, consider the subtraction of the \eqref{OffShellTotalNoetherIdentity} for the old theory from the same expression in the modified theory:
\begin{equation}
    \dd_{\rho}\left(J'^{\rho}-J^{\rho}\right) = \frac{\de S}{\de y_a}\de y_a - \frac{\de S}{\de y'_E}\de y'_E.
    \label{J'JRawRelation}
\end{equation}
Because the change \eqref{ChangeOfVariables} is smooth, one can write a polynomial expansion for $\de y_a$ in terms of $\de y'_E$:
\begin{equation}
    \de y_a=\sum_{i = 0}^W\frac{\dd y_a}{\dd\dd_{\al_1..\al_i}y'_E}\dd_{\al_1..\al_i}\de y'_E,
    \label{YY'Variation}
\end{equation}
and also derive the expression for $\de S/\de y'_a$ in terms of the original equations of motion $\de S/\de y_a$:
\begin{equation}
    \frac{\de S}{\de y'_E} = \sum_{i = 0}^W(-1)^i\dd_{\al_1..\al_i}\left(\frac{\de S}{\de y_a}\frac{\dd y_a}{\dd\dd_{\al_1..\al_i}y'_E}\right).
    \label{YY'EOM}
\end{equation}
By substituting \eqref{YY'Variation} and \eqref{YY'EOM} into \eqref{J'JRawRelation} and then using \eqref{HigherOrderFullDerivativeRelation}
for the resulting formula, one can show the following:
\begin{align}
    \dd_{\rho}\left(J'^{\rho}-J^{\rho}\right) = \dd_{\rho}\sum_{j = 1}^W\sum_{i=0}^{j-1}(-1)^{i+j+1}\dd_{\al_1..\al_{j-i-1}}\left(\frac{\de S}{\de y_a}\frac{\dd y_a}{\dd\dd_{\rho\al_1..\al_{j-1}}y'_E}\right)\dd_{\al_{j-i}..\al_{j-1}}\de y'_E,
    \label{JJ'DivergenceRelation}
\end{align}
Since no equations of motion is required for \eqref{JJ'DivergenceRelation}, it is an identity and hence $J'^{\rho}$ and $J^{\rho}$
obey the following relation:
\begin{equation}
    J'^{\rho} = J^{\rho} + \sum_{j = 1}^W\sum_{i=0}^{j-1}(-1)^{i+j+1}\dd_{\al_1..\al_{j-i-1}}
    \left(\frac{\de S}{\de y_a}\frac{\dd y_a}{\dd\dd_{\rho\al_1..\al_{j-1}}y'_E}\right)\dd_{\al_{j-i}..\al_{j-1}}\de y'_E + I^{\rho},
    \label{J'Constraint}
\end{equation}
where $I^{\rho}$ is a divergence-free term:
\begin{equation}
    \dd_{\rho}I^{\rho} = 0.
\label{J'Constraint2}
\end{equation}
Note that it follows from \eqref{J'Constraint} that $I^{\rho}$ necessarily is a local function of the $y'_E$ and diffeomorphism parameter $\xi^\mu$.
Unfortunately, the explicit formula for this quantity remains unknown.

\section{Examples}\label{primeri}
In this section, we consider several applications of the general formalism described above.
Firstly, we focus on the standard metric description of the GR and then proceed to its modifications.
In all examples we use the standard Einstein-Hilbert action of GR:
\begin{equation}
    S_{gr} = -\frac{1}{\ka}\int d^4x\sqrt{-g}\,R.
    \label{EH}
\end{equation}
For the sake of simplicity will also omit matter contribution by setting $S_m=0$, so $S=S_{gr}$,
nevertheless, all the results below can be easily generalized to the case $S_m\neq 0$.

\subsection{General Relativity}\label{oto}
The relations \eqref{XQuantityDefinition} and \eqref{sp4} give for \eqref{EH} the following:
\begin{align}
    &X^{\rho}{}_{\mu}= \left[V^{\rho}{}_{\mu}\right]_{gr} =-\frac{1}{\ka} \sqrt{-g}\,G^{\rho}{}_{\mu}.
\end{align}
By using the following expressions for $Z^{\rho\nu\be}$ and $Z^{\rho\nu\be\al}$, that are defined in \eqref{ZetNDefinitions}:
\begin{align}
    Z^{\rho\nu\be} =&-\frac{\sqrt{-g}}{2\ka}\left( g^{\rho\nu}g^{\mu\sigma}\Ga^{\be}_{\mu\sigma}-g^{\mu\sigma}\simup{g^{\be\rho}\Ga^{\nu}_{\mu\sigma}}{\rho\nu}+g^{\mu\be}\simup{g^{\al\rho}\Ga^{\nu}_{\mu\al}}{\rho\nu}-g^{\mu\rho}g^{\nu\al}\Ga_{\mu\al}^{\be}\right),
    \label{3IndexZGR}\\
    Z^{\rho\nu\be\al} =&-\frac{\sqrt{-g}}{2\ka}\left( \frac{1}{2}\simup{g^{\al\rho}g^{\be\nu}}{\al\be} - g^{\rho\nu}g^{\be\al}\right),
    \label{4IndexZGR}
\end{align}
the following expression for the superpotential \eqref{sp2} can be established:
\begin{align}
    \Psi_{\text{GR}}{}^{\be\rho}{}_{\al} &=\frac{\sqrt{-g}}{2\ka}\left(2\Ga^{\be}_{\al\ga}g^{\rho\ga}- \Ga^{\mu}_{\mu\al}g^{\rho\be}-\de^{\be}_{\al}\Ga^{\rho}_{\mu\ga}g^{\mu\ga}\right).
    \label{GRKFirst}
\end{align}
One can substitute this expression in the \eqref{GravitySuperpotentialRelation} to obtain an off-shell identity for the
pEMT:
\begin{equation}
\mathcal{T}^{\rho}{}_{\la}=-\frac{1}{\ka} \sqrt{-g}\,G^{\rho}{}_{\la}+\frac{1}{2\ka} \dd_{\be}\left[\sqrt{-g}\left(2\Ga^{\be}_{\la\ga}g^{\rho\ga}- g^{\rho\be}\Ga^{\mu}_{\mu\la}-\de^{\be}_{\la}\Ga^{\rho}_{\mu\ga}g^{\mu\ga}\right)\right].
\label{sp3}
\end{equation}

One can also construct {$L_{\al}{}^{\be\ga\rho}$, described by the formulae \eqref{AntisymmetricSuperpotentialDefinition} and \eqref{sp5}. In that case the expression \eqref{GRKFirst} may be replaced by the antisymmetric quantity $\Psi_{GR}{}^{\be\rho}{}_{\al}+\dd_{\ga}L_{\al}{}^{\be\ga\rho}$. This superpotential is well-known for GR and first was obtained by M{\o}ller \cite{Moller}}:
\begin{align}
    &\Psi_{\text{M}}{}^{\be\rho}{}_{\al} \equiv \frac{1}{\varkappa}\sqrt{-g}\left(g^{\ga\rho}\Gamma^{\be}_{\ga\mu}-g^{\ga\be}\Gamma^{\rho}_{\ga\mu}\right){.}
    \label{MollerSuperpotDef}
\end{align}
{With that in mind, \eqref{sp3} may be rewritten in the following form:}
\begin{align}
    &\mathcal{T}^{\rho}{}_{\la}=-\frac{1}{\ka} \sqrt{-g}\,G^{\rho}{}_{\mu}+\dd_{\be}\left(\Psi_{\text{M}}{}^{\be\rho}{}_{\al}\right).
    \label{GRMollerForm}
\end{align}
This identity reproduces the expected standard form of pEMT for the Einstein field on-shell.

It is interesting that  the main property \eqref{KsuperpotentialProperty} of the superpotential $\Psi_{\text{GR}}{}^{\be\rho}{}_{\al}$
is satisfied identically.
For M{\o}ller expression this statement trivially follows from its antisymmetry and for the $\Psi_{\text{GR}}{}^{\be\rho}{}_{\al}$
it can be seen from the property \eqref{sp5} of the $L_{\al}{}^{\be\gamma\rho}$.

\subsection{Palatini formalism}
By Palatini (or Hilbert-Palatini) formalism one usually means {the description of gravity with the symmetric connection and spacetime metric as the independent variables. The action in such approach is taken in the standard Einstein-Hilbert form \eqref{EH}.}
Despite the name of the approach, for the first time it was probably proposed by Einstein in \cite{ein_conn}, for the detailed
discussion of the topic see \cite{ferraris_pal}.
By itself, Palatini formalism does not provide the real modification of GR; however,
it is a useful example of the theory with independent non-tensor variables.

Quantities $H_{(k)\la}{}^{\al}{}_{\mu\nu}{}^{\al_1...\al_k}$ for the connection $\Ga^{\al}_{\mu\nu}$ are the following (recall
that symmetrization brackets introduced after \eqref{OmegaThirdDefinition} is not conventional):
\begin{equation}
    H_{(0)\la}{}^{\al}{}_{\mu\nu} = \dd_{\la}\Ga^{\al}_{\mu\nu},\;\;\;\;
    H_{(1)\la}{}^{\al}{}_{\mu\nu}{}^{\rho} = \simdown{\de^{\rho}_{\mu}\Ga^{\al}_{\la\nu}}{\mu\nu}-\de^{\al}_{\la}\Ga^{\rho}_{\mu\nu},\;\;\;
    H_{(2)\la}{}^{\al}{}_{\mu\nu}{}^{\rho\nu} = \frac{1}{2}\de^{\al}_{\la}\simdown{\de^{\rho}_{\mu}\de^{\nu}_{\nu}}{\mu\nu}.
    \label{Hilbert-PalatiniHQuantities}
\end{equation}
By using \eqref{sp2} and \eqref{Hilbert-PalatiniHQuantities} it can be shown that $\Psi^{\be\rho}{}_{\al}$ for the Palatini
formalism coincides with the $\Psi_{\text{GR}}{}^{\be\rho}{}_{\al}$ defined in \eqref{GRKFirst}.
Thus, the pEMT \eqref{GravitySuperpotentialRelation} for the considered theory has the form:
\begin{align}
\nonumber
    \mathcal{T}^{\rho}{}_{\la}=-&\frac{1}{\ka} \sqrt{-g}\,G^{\rho}{}_{\la}+\\+&
    \frac{1}{2\ka}\sqrt{-g}\antisimup{g^{\mu\nu}C^{\al}{}_{\al\ga}-g^{\sigma\nu}C^{\mu}{}_{\sigma\al}\de^{\al}_{\ga}}{\al\nu} \left(\simdown{\de^{\rho}_{\mu}\Ga^{\ga}_{\nu\la}}{\mu\nu}-\de^{\ga}_{\la}\Ga^{\rho}_{\mu\nu}\right)+\dd_{\be}\Psi_{\text{GR}}{}^{\be\rho}{}_{\la},
\end{align}
where $C^{\mu}{}_{\sigma\al}\equiv \Ga^{\mu}_{\sigma\al}-\overset{\text{o}}{\Ga}{}^{\mu}_{\sigma\al}$ denotes the difference
between total connection $\Ga^{\mu}_{\sigma\al}$ and the Levi-Civita connection $\overset{\text{o}}{\Ga}{}^{\mu}_{\sigma\al}$, and
$[..]_{\mu\nu}$ denotes antisymmetric part (without normalization) of the quantity in the brackets with
respect to the given indices.

It is worth noting that, in contrast with GR, {for} Palatini formalism, one cannot state that the superpotential property \eqref{KsuperpotentialProperty}
of $\Psi^{\be\rho}{}_{\al}$ is satisfied identically.
Indeed, if one considers \eqref{KDecompositionReccurentChain1} with $k = 2$ and takes into account
that the equations of motion arising from the variation principle with respect
to $\Ga^{\mu}_{\al\nu}$ have the form $C^{\mu}{}_{\sigma\al}=0${. It} can be easily checked that the r.h.s of \eqref{KDecompositionReccurentChain1} with $k = 2$
should be proportional to the $C^{\mu}{}_{\sigma\al}$:
\begin{equation}
    \dd_{\rho}K_{(2)\la}{}^{\rho\pp\sigma}+\frac{1}{2}\simup{K_{(1)\la}{}^{\pp\sigma}}{\pp\sigma} = -\frac{1}{2\ka}\sqrt{-g}\antisimup{g^{\pp\sigma}C^{\al}{}_{\al\la}-\frac{1}{2}\simup{g^{\ga\pp}C^{\sigma}{}_{\ga\al}\de^{\al}_{\la}}{\pp\sigma}}{\al\nu}.
\end{equation}
It immediately implies that \eqref{KsuperpotentialProperty} is satisfied only when the equations of motion corresponding to $\Gamma^{\rho}_{\mu\nu}$
are satisfied, namely when the connection is reduced to the Christoffel symbols.
Obviously enough, when all the equations of motion are satisfied, the expression for pEMT in Palatini formalism also
coincides with that one for GR.

\subsection{Disformal transformations and mimetic gravity}\label{MimeticGravitySec}
For modification of GR with the action \eqref{EH} one can use well-known change of the independent variables that is
called \textit{disformal transformation} \cite{DisformalBekenstein}:
\begin{equation}
    g_{\mu\nu} = A(Q,\sigma)\bar{g}_{\mu\nu} + B(Q,\sigma)(\dd_{\mu}\sigma)(\dd_{\nu}\sigma),
    \label{DisformalChange}
\end{equation}
where $\bar{g}_{\mu\nu}$ is an auxiliary metric, $\sigma$ is a scalar field, $Q \equiv \bar{g}^{\mu\nu}(\dd_{\mu}\sigma)(\dd_{\nu}\sigma)$.
From \eqref{J'Constraint} it can be shown that for the changes like \eqref{DisformalChange} the following relation holds:
\begin{equation}
    J'^{\rho} = J^{\rho}-\frac{1}{\ka}\sqrt{-g}G^{\mu\nu}\left((A'_{\sigma}\bar{g}_{\mu\nu}+B'_{\sigma}\dd_{\mu}\sigma\dd_{\nu}\sigma)\bar{g}^{\rho\al}\dd_{\al}\sigma+\de^{\rho}_{\mu}B\dd_{\nu}\sigma\right)\xi^{\al}\dd_{\al}\sigma+I^{\rho},
\label{spn1}
\end{equation}
and one should expect the superpotential of the modified theory to differ from the original one \eqref{GRKFirst} only by the
the divergence-free term arising from $I^{\rho}${. This statement follows from the observation, that second term in the r.h.s depends on $\xi^{\al}$ without any
derivatives and hence does not contribute to the $K_{(1)\al}{}^{\be\rho}$. It also can be easily verified}
 by the direct calculation of the \eqref{sp2},
which gives the following off-shell expression for the pEMT:
\begin{equation}
    \mathcal{T}^{\rho}{}_{\la} = V^{\rho}{}_{\la} +\dd_{\be}\left(\Psi_{\text{GR}}{}^{\be\rho}{}_{\mu}+\Psi_{\text{add}}{}^{\be\rho}{}_{\mu}\right),
\label{spn2}
\end{equation}
where
\begin{align}
    &V^{\rho}{}_{\la} = \frac{1}{\ka}\sqrt{-g}\left(AG^{\rho}{}_{\la} - G^{\al\be}\left(A'_Q\bar{g}_{\al\be}+B'_Q\dd_{\al}\sigma\dd_{\be}\sigma\right)\dd^{\rho}\sigma\dd_{\la}\sigma\right),
    \label{VForDisformal}\\
    \nonumber
    &\Psi_{\text{add}}{}^{\be\rho}{}_{\mu} = -
    \Bigg[Z^{\pp\sigma\al}\frac{\dd\dd_{\al}g_{\pp\sigma}}{\dd\dd_{\be}\bar{g}_{\rho\nu}}\bar{g}_{\nu\mu}-
    Z^{\rho}{}_{\mu}{}^{\be}+Z^{\pp\sigma\ep\omega}\frac{\dd\dd_{\ep\omega}g_{\pp\sigma}}{\dd\dd_{\be}\bar{g}_{\rho\nu}}\bar{g}_{\mu\nu}+
    2Z^{\pp\sigma\ep\omega}\frac{\dd\dd_{\ep\omega}g_{\pp\sigma}}{\dd\dd_{\be\al}\bar{g}_{\rho\nu}}\dd_{\al}\bar{g}_{\nu\mu}-\\
    &\hspace{4cm}-
    2Z^{\rho\nu\be\al}\dd_{\al}g_{\mu\nu}-\frac{4}{3}\dd_{\al}\left(Z^{\be\nu\al\rho}g_{\mu\nu}+
    Z^{\pp\nu\ep\omega}\frac{\dd\dd_{\ep\omega}g_{\pp\nu}}{\dd\dd_{\be\al}\bar{g}_{\ga\rho}}\bar{g}_{\ga\mu}\right)\Bigg]^{\rho\be},
    \label{RawDisformalSuperpotential}
\end{align}
and $Z^{\pp\sigma\al}$, $Z^{\pp\sigma\ep\omega}$ are defined in \eqref{3IndexZGR} and \eqref{4IndexZGR}.
Here we used \eqref{3IndexZGR}, \eqref{4IndexZGR} and \eqref{GravitySuperpotentialRelation}.
We can again replace $\Psi_{\text{GR}}{}^{\be\rho}{}_{\mu}$ with the M{\o}ller superpotential \eqref{MollerSuperpotDef} (see section~\ref{oto}),
which leads to the simplified expression for pEMT through the antisymmetric superpotential:
\begin{align}
    \mathcal{T}^{\rho}{}_{\la} = V^{\rho}{}_{\la} +\dd_{\be}\left(\Psi_{\text{M}}{}^{\be\rho}{}_{\mu}+\Psi_{\text{add}}{}^{\be\rho}{}_{\mu}\right).
    \label{AntisymmetricDisformalSuperpotential}
\end{align}

As it was noted at the end of the section~\ref{SuperpotentialStructureSec}, if $H_{(1)a\mu}{}^{\rho} = 0$ for any independent variable, the quantity
$V^{\rho}{}_{\la}$ does not depend on the equations of motion that arise {from} the variational principle for this variable.
Metric theories with the disformal change provide an example exactly of this kind: {in} this case, $\sigma$ is the mentioned field.
As a result, it can be seen from the \eqref{VForDisformal} for $V^{\rho}{}_{\la}$ that the equations of motion, which arise from
the variation with respect to $\bar{g}_{\mu\nu}$ are sufficient for the continuity equations for the $\mathcal{T}^{\rho}{}_{\la}$
to be satisfied.

One of the most popular choice of the disformal change is so-called mimetic gravity proposed in the paper \cite{mukhanov}.
In that case the change takes the simple form:
\begin{equation}
    g_{\mu\nu} = \bar{g}_{\mu\nu}\bar{g}^{\al\be}(\dd_{\al}\sigma)(\dd_{\be}\sigma),
    \label{MimeticChange}
\end{equation}
which corresponds to the choice $A(Q,\sigma)=Q$, $B(Q,\sigma)=0$ in \eqref{DisformalChange}.
It is crucial that this particular case is somewhat singular because the change \eqref{MimeticChange} is not invertible  in contrast with the regular case of \eqref{DisformalChange}
\cite{arXiv1311.3111}.
For mimetic gravity, additional contribution to the superpotential is simplified to the expression:
\begin{align}
    \Psi_{\text{add}}{}^{\be\rho}{}_{\mu} = -\bigg[Z^{\pp}{}_{\pp}{}^{\rho}\dd^{\be}\sigma\dd_{\mu}\sigma+
    2Z^{\pp\sigma\ep\rho}\dd_{\ep}\left(g_{\pp\sigma}\dd^{\be}\sigma\dd_{\mu}\sigma\right)-\frac{4}{3}\dd_{\chi}\left(Z^{\pp}{}_{\pp}{}^{\rho\chi}
    \dd^{\be}\sigma\dd_{\mu}\sigma\right)\bigg]^{\rho\be}.
    \label{MimeticSuperpotential}
\end{align}
Substituting this formula in \eqref{AntisymmetricDisformalSuperpotential} gives the expression for canonical pEMT in mimetic gravity.

Apart from the form of the mimetic gravity induced solely by the change \eqref{MimeticChange} in action \eqref{EH}, several
generalizations with the potential for $\sigma$ are often considered (see, for example, \cite{mukhanov2014}).
In cases when
this potential does not contain derivatives, the formula \eqref{MimeticSuperpotential} still holds because \eqref{OmegaFirstDefinition}
(and hence the superpotential \eqref{sp2}) does not depend on $Z_{(\sigma)}\equiv \dd L/\dd\sigma$ (see \eqref{ZetNDefinitions}).

The change \eqref{MimeticChange} extends the dynamic of GR in such way that the modified theory is equivalent to the GR with the
additional "mimetic"{} matter.
{Since the new matter originates from the gravity only, this behaviour allows one to approach certain problems of the modern cosmology from a new angle. For instance, one may}
identify this mimetic matter with cold dark matter, which interacts only through gravity and thus may be described as
pure gravitational effect (see \cite{mimetic-review17} and its references for the current state of the topic).
The additional
extension of dynamics can be achieved {in two ways. The first possibility is to just change the action  -- for example, by adding a potential to the scalar field. Another variant is to modify the change \eqref{MimeticChange} itself (see, for example, the change proposed in \cite{statja48} for which
dark matter has a non-potential flow)}.

It should be noted that theory with change \eqref{MimeticChange} in \eqref{EH} can be equivalently reformulated as GR with
the additional action term by adding the Lagrange multiplier \cite{Golovnev201439}.
It can be checked that the superpotential
for this reformulated theory coincides with \eqref{GRKFirst}, and hence superpotential and pEMT again can be reduced to the M{\o}ller superpotential
\eqref{MollerSuperpotDef} and \eqref{GRMollerForm} respectively.

\subsection{Regge-Teitelboim embedding gravity}\label{Regge-TeitelboimSec}
Proposed by Regge and Teitelboim \cite{regge}, embedding gravity (or "embedding theory"{}) is based on the idea of
considering four-dimensional GR manifold as an embedded surface $\mathcal{M}$ in a flat spacetime of higher dimension $d$.
Action is chosen as usual \eqref{EH} with the induced metric on $\mathcal{M}$ playing role of the change of variables in action:
\begin{equation}
    g_{\mu\nu} = (\dd_{\mu}y^a)(\dd_{\nu}y_a).
    \label{EmbeddingChangeOfVariables}
\end{equation}
Here $y^a$ is an embedding function, indices $a,b,..$ in this subsection denotes the tensor representations of the symmetry
group $SO(1,d-1)$ of the ambient space flat metric.
It is essential that \eqref{EmbeddingChangeOfVariables} requires the isometricity of the proper embedding.
The minimal dimension of the ambient space for the embedding of the smooth four-dimensional Lorentzian manifold, as it was pointed out
in \cite{fridman61}, is $d=10$.
After the paper \cite{regge} the framework of embedding gravity was thoroughly studied in \cite{deser}. {Since then} these ideas were often used in the various investigations, including researches on the quantum gravity (see, for example, \cite{pavsic85,tapia,estabrook1999,davkar,statja18,rojas09,faddeev,statja26, statja33}).
In the full analogy to the change \eqref{MimeticChange} the embedding change \eqref{EmbeddingChangeOfVariables} extends the dynamics
of the original GR{. Indeed, just like the mimetic gravity, the modified theory appears to be equivalent to GR with an extra matter}.
Interestingly enough, the number of degrees of freedom corresponding to the extra matter in the resulting theory seems to be big enough to describe dark matter without additional complications \cite{statja51}.

Due to the scalar origin of the independent variables (the embedding function $y^a$) of the embedding gravity
the formula \eqref{GravitySuperpotentialRelation}
reduces (see the end of section~\ref{SuperpotentialStructureSec}) to the form \eqref{ScalarOffShellContinuityEquation} even without any
equations of motion.
The direct calculation based on the \eqref{sp2} allows one to write \eqref{ScalarOffShellContinuityEquation} in the form:
\begin{equation}
    \mathcal{T}^{\rho}{}_{\la} = \dd_{\be}\left(\Psi_{\text{GR}}{}^{\be\rho}{}_{\la}+\Psi_{\text{add}}{}^{\be\rho}{}_{\la}\right),
    \label{embeddingPEMT}
\end{equation}
where
\begin{equation}
\Psi_{\text{add}}{}^{\be\rho}{}_{\mu} = \left[Z^{\rho}{}_{\mu}{}^{\be}+2Z^{\rho\nu\be\al}\dd_{\al}g_{\mu\nu}+
\frac{4}{3}\dd_{\al}\left(Z^{\be\nu\rho\al}g_{\mu\nu}\right)\right]^{\be\rho},
\label{EmbeddingSuperpotential}
\end{equation}
and quantities $Z^{\rho\mu\be}$, $Z^{\rho\nu\be\al}$ are defined in \eqref{3IndexZGR}, \eqref{4IndexZGR}.
By replacing $\Psi_{\text{GR}}{}^{\be\rho}{}_{\mu}$ {with} the M{\o}ller superpotential \eqref{MollerSuperpotDef} likewise
it has been done above, the canonical pEMT finally takes the {antisymmetric} form:
\begin{equation}
    \mathcal{T}^{\rho}{}_{\la} = \dd_{\be}\left(\Psi_{\text{M}}{}^{\be\rho}{}_{\la}+\Psi_{\text{add}}{}^{\be\rho}{}_{\la}\right).
    \label{EmbeddingPseudotensor}
\end{equation}
Thus, the local conservation law \eqref{ScalarOffShellContinuityEquation2} for pEMT {\eqref{embeddingPEMT}} is satisfied without any equations of motion, and the pEMT itself can be identically rewritten through the superpotential.
Iterestingly enough, these properties also hold in the case $S_m \neq 0$ with the  satisfied equations of motion of matter.
In a way, the embedding gravity can be treated as the "extremal case" of the theory whose metric components are isolated by a change like \eqref{MimeticChange}.
Mimetic change isolates the conformal mode of the metric into the kinetic term of the scalar field, which
makes the corresponding to that scalar field equation of motion unnecessary for the pEMT continuity equation{. In cotrast, for the embedding gravity, all degrees of freedom find themselves in that position}.

An another interesting aspect of embedding gravity is worth mentioning, namely the geometrical interpretation
of the second Noether theorem (the identities \eqref{MetricEMTCovariantConservationGeneralization}) in this case.
At the beginning of the section~\ref{SuperpotentialStructureSec}, we noted that the second Noether theorem
\eqref{MetricEMTCovariantConservationGeneralization}
establishes the relation between the equations of motion arising from the general covariance of the action.
For theory with the disformal change \eqref{DisformalChange}, this relation comes down to the fact that the equations
of motion corresponding to the $\sigma$ are automatically satisfied for any solution of the equation of motion corresponding to $\bar{g}_{\mu\nu}$.
It makes the absence of requirement $\de S/\de \sigma = 0$ in the continuity equation for the pEMT \eqref{spn2}
(by taking \eqref{VForDisformal} into an account) quite natural.
For the embedding theory the identity \eqref{MetricEMTCovariantConservationGeneralization} has the following form:
\begin{equation}
\frac{\de S}{\de y^a}H_{(0)}{}^a{}_\mu = 0,\qquad
H_{(0)}{}^a{}_\mu= \dd_{\mu}y^a.
\label{ReggeTeitelboimIdentity}
\end{equation}
It is obvious that the set $\left\{\dd_{\mu}y^a, \mu = 0\dots3\right\}$ forms a basis in the tangent space of the surface $\mathcal{M}${. So
the formula} \eqref{ReggeTeitelboimIdentity} geometrically means that the tangent ({with respect to} $\mathcal{M}$) part
of the equations of motion for the embedding gravity holds identically, which is a well-known fact for the Regge-Teitelboim
theory equations (see, for example, \cite{statja18}).
The relation \eqref{ReggeTeitelboimIdentity} also makes \eqref{KDecompositionReccurentChain0} consistent with the local conservation law
\eqref{ScalarOffShellContinuityEquation2} for the pEMT \eqref{EmbeddingPseudotensor}.

\section{Concluding remarks}\label{ResultsAndDiscussionSec}
Despite the strong connection with the Noether procedure described in formulae \eqref{LagrangianTransfLaw}-\eqref{KDecompositionReccurentChain2}, the
choice of $J^{\rho}(\xi)$ precisely in the form \eqref{JDefinition}
for construction of the recurrent chain \eqref{KDecompositionReccurentChain0}-\eqref{KDecompositionReccurentChain2} is not unique.
Indeed, for theories discussed in the section~\ref{ChangeOfVariablesSection} with the change of the independent variables \eqref{ChangeOfVariables} in action
one may start from the identities \eqref{OffShellTotalNoetherIdentity} for the original theory instead of the new ones
and rewrite them in form:
\begin{equation}
    \dd_{\rho}J^{\rho} -\left(\frac{\de S}{\de y_a}\de y_a - \frac{\de S}{\de y'_E}\de y'_E\right) = \frac{\de S}{\de y'_E}\de y'_E
    \label{J'-likeCurrentConstruction}
\end{equation}
(for simplicity we set $S_m = 0$, however, the general case actually does not bring anything new).
The second term can be further brought to form of the full divergence by using the same logic which was used
in simplifying of the r.h.s. of the relation \eqref{J'JRawRelation}.
The resulting relation will look like \eqref{OffShellTotalNoetherIdentity} for the modified theory,
and the final expression for the new definition of $J^{\rho}$ will be equal
to the r.h.s of \eqref{J'Constraint} with $I^{\rho} = 0$, which may differ from the direct calculation based on the formulae
from section~\ref{GeneralSuperpotentialSec}.
Another example of the alternative procedure of deriving the densities for conserved charges is provided by the calculation of the
superpotential for Regge-Teitelboim embedding gravity in \cite{statja46}.
In that paper, certain identities for the original theory (namely, GR) are used, and in the end, the obtained
superpotential $\Psi^{\be\rho}{}_{\mu}$ coincided with the Moller one, and hence it differs from the expression in section~\ref{Regge-TeitelboimSec}
by the addition \eqref{EmbeddingSuperpotential}.
Nevertheless, it can be shown that $J^{\rho}$ with the definition from section~\ref{GeneralSuperpotentialSec} in
case of the first-order derivative theory is closely connected with the Ward identities arising in the quantization of the gauge theories,
particularly for the generally covariant ones \cite{AveryNoetherWard}.
{We expect this property to hold in the general case. In this regard, it is interesting how the obtained formula \eqref{J'Constraint} is reflected in the general form of Ward identities for BMS symmetry (see, for example, \cite{BMSOriginal,HawkingBMS}) for theories with changes \eqref{ChangeOfVariables}. This topic goes beyond the discussion in the current paper and is a subject of further studies.}

{Another application of the results obtained is to use the formulae \eqref{J'Constraint},\eqref{EmbeddingSuperpotential} and \eqref{AntisymmetricDisformalSuperpotential} for the analysis of cosmological perturbations in some GR modifications. As it is shown in the papers \cite{linden-bell, petrov2}, the conservation laws may be used for derivation of the integral constrains, introduced originally in \cite{Traschen} for FLRW metrics. The existence of such constraints significantly reduces the Sachs-Wolfe effect \cite{TraschenErdley} on the mean value of angular fluctuations of the cosmic
background radiation. At the moment, this question is scarcely been explored for theories with changes \eqref{ChangeOfVariables}, and needs to be developed further.}

Throughout the paper, the formula \eqref{KDecomposition} was used only for the analysis of the canonical pEMT, which corresponds to the
conserved energy-momentum vector \eqref{Energy-MomentumConservation}.
This identity can also be used to calculate the pseudotensor $U^{\mu\al\be}$ which defines the density of the total angular momentum tensor $M^{\al\be}$
and the superpotential for it.
To do it, one should again consider a diffeomorphism $x^{\mu}\rightarrow x^{\mu}+\xi^{\mu}$ and write down the
infinitesimal diffeomorphism parameter in the special non-covariant form:
\begin{equation}
\xi^{\mu}(x) = \omega_{\al\be}(x)\hat{M}^{\mu\al\be}(x),\qquad
\hat{M}^{\mu\al\be}(x) \equiv (x^{\al}\eta^{\mu\be} - x^{\be}\eta^{\mu\al}),
\end{equation}
where $\eta^{\mu\be}$ denotes metric of the Minkowski space, and $\omega_{\al\be}(x)$ -- arbitrary function with the
antisymmetry property $\omega_{\al\be} = - \omega_{\be\al}$.
If one substitute this expression for $\xi^\mu$ in \eqref{KDecomposition} and then use the arbitrariness of
$\omega_{\al\be}$ and its derivatives at one point, the new chain can be derived, similar to the one defined in
\eqref{KDecompositionReccurentChain0}-\eqref{KDecompositionReccurentChain2}.
This chain leads to the equations for the quantities associated with angular momentum tensor,
analogous to the \eqref{SuperpotentialAnalysisResult} and \eqref{Energy-MomentumConservation} with
the recalculated quantities $K_{(1)a\mu}{}^{\al_1..\al_k}$.
Though $U^{\mu\al\be}$ is not tensor with respect to the diffeomorphisms, it can be easily {verified} that the corresponding angular momentum
$M^{\al\be}$ is still a tensor with respect to the Lorentz group. { The reason for this is obviously one's} ability to treat $M^{\al\be}$ as a surface term
like it is done for the energy-momentum vector \eqref{Energy-MomentumConservation}.

{\bf Acknowledgements.}
The authors are grateful to A.~Sheykin for useful discussion.
The work is supported by RFBR Grant No.~20-01-00081.

%\newcommand{\eprint}[1]{\href{http://arxiv.org/abs/#1}{\texttt{#1}}}
%\bibliographystyle{../../my3}
%\bibliography{../../paston-grav-e}
%\end{document}

\end{document}